\documentclass[agupp]{aguplus}
\usepackage{graphics}

\renewcommand{\arraystretch}{1.2}
\lefthead{Baumert}
\righthead{Universal constants and equations}
\sectionnumbers
\renewcommand{\fboxrule}{0pt}

\authoraddr{H. Z. Baumert, IAMARIS, Bei den M\"uhren 69A,\\
D-20457 Hamburg, Germany (baumert@iamaris.org)}

\slugcomment{\it Submitted as v2 to {\bf arXiv.org} \today
}

\usepackage{graphicx}
\graphicspath{    {converted_graphics/}{/}
}
\begin{document}
\bibliographystyle{ametsoc}

\def\lesssim{\mathrel{\hbox{\rlap{\hbox{\lower0.45em\hbox{$\sim$}}}\hbox{$<$}}}}
\def\gtrsim{\mathrel{\hbox{\rlap{\hbox{\lower0.45em\hbox{$\sim$}}}\hbox{$>$}}}}
\def\gae{\mathrel{\hbox{\rlap{\hbox{\lower0.45em\hbox{$\sim$}}}\hbox{$>$}}}}
\def\lae{\mathrel{\hbox{\rlap{\hbox{\lower0.45em\hbox{$\sim$}}}\hbox{$<$}}}}
\title{
\vspace{0.75cm}
\vspace{0.75cm}
Universal constants  and \\ equations of turbulent motion}

\author{by Helmut Z.\ Baumert,}
\affil{Institute for Applied Marine and Limnic Studies (IAMARIS e.V.), Hamburg, Germany}


\begin{abstract}
In the spirit of Prandtl's [1926] conjecture, \nocite{prandtl26} for turbulence at $Re\rightarrow \infty$ we present an analogy with the kinetic theory of gases, 
with dipoles made of 
quasi-rigid and ``dressed'' vortex tubes as frictionless, incompressible but deformable quasi-particles. Their movements are governed by Helmholtz' elementary 
vortex rules applied locally. A contact interaction or ``collision'' leads either to random scatter of a trajectory or to the formation of two likewise 
rotating, fundamentally unstable whirls forming a dissipative patch slowly rotating around its center of mass which is almost at rest. 
This approach predicts  von Karman's constant  as $\kappa=1/\sqrt{2\pi }\simeq 0.399$ and the spatio-temporal dynamics 
of energy-containing time and length scales controlling turbulent mixing  \cite[][]{baumert09} in agreement with observations. 
A link to turbulence spectra was missing so far. In the present paper it is shown that the above image of random vortex-dipole 
movements is compatible with Kolmogorov's turbulence 
spectra if dissipative patches, beginning as two $likewise$ rotating eddies, evolve locally into a space-filling bearing in the sense of  
\cite{herrmann90}, i.e.\ into an ``Apollonian gear'' consisting of incompressible and flexibly deformable vortex tubes which 
are frictionless, excepting the dissipative scale of size zero.

\indent{    }
For steady and locally homogeneous conditions our approach predicts the pre-factor in the three-dimensional Eulerian wavenumber spectrum,
$E(k)= \alpha_1\, \varepsilon^{2/3}\, k^{-5/3} $, as $\alpha_1 = \frac{1}{3}(4\,\pi)^{2/3}\simeq 1.802$, and  in the Lagrangian frequency spectrum, 
$E(\omega)=\beta_1\,\varepsilon \,\omega ^{-2}$, as $\beta_1 =2$. The unique values for $\alpha_1, \beta_1$ and $\kappa$ are situated well within 
the broad  scatter range of observational, experimental and approximative results. Our derivations rest on geometry, methods from many-particle physics,
and on elementary conservation laws. 
\end{abstract}

\nocite{imbergerboashash86,davidson04,
landaulifshitz_eng87,lugt79,
thrin2010,sreenivasan95}

\vspace{5cm}


\section{Introduction\label{intro}}

\noindent
In the present paper we show that fluid turbulence can be understood  
in an idealised sense as a statistical many-body ensemble -- a tangle 
of vortex tubes taken as \textit{discrete} particles. We follow an early 
conjecture by Ludwig \cite{prandtl26} 
who discussed an analogy between molecular diffusion and turbulence. 
He related his mixing length\footnote{According to \cite{hinze59} also G. I. Taylor made early use
of the notion \textit{mixing length}.} or \textit{Mischungsweg} with the mean-free path 
of kinetic gas theory and considered his fluid elements or fluid lumps (\textit{Fl{\"u}ssigkeitsballen})
of locally nearly same size as relatives of  gas molecules. He further assumed his 
mixing length to scale with the ``diameter'' of his fluid elements. 

Although purely heuristic, his concept became popular in the years before WW2.  
The question how to compute the details could not be answered without reference to
measurements. After WW2 the gas-kinetic analogy found thus strong criticism  from the continuous-image side
\cite[see Introduction in][]{batchelor53}. This was plausible because the gas analogy represents a \textit{discrete} concept that, 
if useful, would label a Copernicanian turn or 
a paradigm shift in our grown view
of turbulence as an exclusive matter of \textit{continuum} mechanics.

In this sitution Werner \cite{albring81}, in the footsteps of 
Prandtl, 
explicitely challenged the continuous paradigm of \cite{reynolds1895}, 
\cite{kellerfriedmann24}, \cite{taylor35}, \cite{batchelor53} and their many followers, 
when he posed the fundamental question:
\begin{quotation}\noindent
	Can  the Navier-Stokes equation be used to calculate turbulent flows?
\end{quotation} 
His doubts were based on the feeling that the continuous image of RANS does
possibly override features of most elementary vortex interactions at small scales. 

In analogy to Albring one might ask: Can we deduce a rose's blossom from the periodic system 
of chemical elements? Intuitively we answer with \textit{no}. However, this answer is justified 
because generally ``more is different'' \cite[]{anderson72}. 

Looking in this sense at Albring's above question we may add that the  prediction horizon of the Navier-Stokes equation (NSE)
is strongly limited by a series of higher-order non-equilibrium phase transitions when a growing Reynolds 
number goes through a number of critical values. In these super-critical regions 
the sensitivity against initial conditions becomes relevant and leads to the famous butterfly effect \cite[][]{lorenz63}. 
This is closely related with irreversibility. 
At least in the limit of vanishing viscosity the initial-value problem for 
NSE looks like a \textit{reversible} one. But we know that 
turbulence at $Re=\infty$ is an \textit{irreversible} process.

In principle this problem was already known by Poincare and others as an aspect of the three-body 
problem of celestial mechanics, but  in a fluid-mechanical context it has been demonstrated only after WW2 by \cite{lorenz60,lorenz63}. 
Without going into details of the onset of turbulence and non-equilibrium phase transitions  in hydrodynamics and their 
reflections in various branches of pure mathematics, together with Prandtl and Albring we hypothesize that 
NSE is \textit{not} sufficient to understand the secrets of turbulence or to even 
 predict turbulent flows, in particular not 
in the limit $Re\rightarrow\infty$. Some readers might even go a step further and
argue that higher-order elements of the Friedman-Keller expansions of NSE,
e.g.\ the second and third-order turbulence closures discussed in 
\cite{voropaeva07} and the literature quoted therein, as well as the various ``corrections'' of those closures populating turbulence theory,  
are modern analogues of the epicycles of geocentric times. But this would exceed the limits of the present report.
In this respect we better refer to the agreeable picture of  today's status of turbulence science drawn by \cite{davidson04} in the 
preface of his book where he mentions even ``religious wars\dots between the different camps'' of turbulence theory.

There are not  many physical phenomena resisting theoreticians so long like turbulence. 
In the past it was mainly supra-conductivity which took about 30 years \cite[][]{feynman63b}. 
Today similarly obstinate problems are dark matter, dark energy, and super-symmetry which
have also now an age of about up to 40 years. But turbulence waits still much longer for redemption 
and remained particularly as \textit{the} very last unsolved enigma of \textit{classical} physics.
\nocite{pullinsaffman98}

Below we develop an asymptotically invariant alternative to RANS 
and higher-order closures, which is  not primarily based on NSE 
but does not violate NSE either. 

Although the theory of many-particle physics offers a huge reservoir of potentially helpful methods and tools, 
a closer inspection reveals that the number of \textit{directly} applicable tools is less impressive. 
Neither Liouville theorem, ergodic hypothesis or  Hamilton formalism nor other concepts 
for thermodynamic equilibrium are applicable in their classical forms. 
Turbulence is essentially an open-system, 
thermodymically non-equilibrium phenomenon. In  best cases we have a steady state or \textit{Flie\ss gleichgewicht} 
in the sense of \cite{bertalanffy53}, \cite{glansdorffprigogine71} and \cite{haken78,haken1983}.
However, the concepts of Ising about particle dressing, quasi-particles and renormalization -- in the broader sense
of \cite{dresden93} and \cite{mccomb2004} -- 
appeared as essential guidelines in the slow evolution of our thoughts.

Surely, turbulence can  successfully be attacked from more than one side\footnote{Theories generally  compromise 
portrait and design aspects and in some case mathematically different images of turbulence  will 
show up eventually as fully equivalent in their physical predictions, 
like Heisenberg's 
matrix mechanics and Schr{\" o}dinger's wave mechanics.}. 
Whereas the continuous image has led to a number of important results, for instance RANS and the broad spectrum of different heuristic closure schemes
discussed e.g.\ by \cite{wilcox06}, without additional phenomenological input 
it gave neither unique values of the universal constants  of turbulent motion like von-Karman's and 
Kolmogorov's spectral constants nor a  \textit{closed} image of turbulent flows.
This has led to the following view shared by many theoreticians
\cite[][p. 173]{landaulifshitz_eng87}:
\begin{quotation}\dots 
and $\kappa$ (is) a numerical constant, the \textit{von Karman constant}, whose value cannot be calculated theoretically 
and must be determined experimentally. It is found to be $\kappa = 0.4$.
\end{quotation}
In this respect one can actually be more optimistic. Frictionless turbulence is governed 
in a weak sense by the Euler equation, a special case of NSE. The Euler equation together with 
the conservation of mass represent pure ``inert geometry''.
We can thus expect that the universal constants of turbulent motion are ruled by certain geometrical 
constants, e.g.\ by irrational numbers like $\sqrt{2}$ and/or transcendental numbers 
like{\mathversion{bold} $\pi$}. Indeed, this is possible. In a precursor  study based on the particle picture used also below,
von Karman's constant could be derived as $\kappa =1\left/\sqrt{2\pi }\right.\simeq  0.399$ \cite[]{baumert09}. 
This interesting result led to the conjecture that the particle concept can give also insight into spectral aspects of turbulence.


Below we make explicit use of theoretico-physical thinking, which differs from other scientific activities like
computer modelling and mathemtical physics, from applied mathematics, and from the measuring and observational disciplines.

Computer modelling and measurements have in common that they deal with really existing, 
\textit{finite} systems and data, in particular with \textit{real} and \textit{integer} numbers, with variables of \textit{finite} size. 

Theoretical physics, however, deals with ``things'' which never did, which do not and never will exist: e.g.\ with 
point masses, homogeneous continua, idealized vortex tubes, linear waves, frictionless fluids, 
plane and impenetrable walls, infinitely small or large variables.
A further building block of theoretical
physics is an often unspoken principle which we apply also here: it ``dictates that,
all things being equal, one goes for the simplest possibility--a rule that has worked
remarkably well.'' \cite[][p. 85, see also Chandrasekhar, 1979]{zee86}. Sometimes
this principle is called Occam's razor.
\nocite{chandrasekhar79}

Mathematics also deals with similar non-existing objects, but their relations to the real world
are outside its responsibility. It is the responsibility of theoretical physics  to relate idealized 
thought systems with the real world. The following report concentrates solely 
on the latter aspects.

With one exception, the following text contains  no ``fancy shmancy mathematics''. 
It is the relation between equations (\ref{langevin3}), (\ref{langevin4}) and (\ref{langevin5}) that
would need some time to be derived from scratch. Furtunately, this has already been done
by other authors many years ago and went into the textbooks on stochastic-dynamic systems 
\cite[e.g.][]{kraichnan68,haken78,haken1983,stratonovich92,stratonovich94} to which we refer
the theoretically interested reader. We further suggest to begin  first with \cite{baumert09} because 
it is actually the basis of the present report.

Everywhere the pronomen ``we'' is used in this text,
 it means the two of us, the dear reader and the author. 


\section{Particles}\label{particles}
\noindent
If we take Prandtl's discrete particle concept 
seriously but not literally, 
then a number of questions need to be posed -- and answered. A first question to be addressed is the following.

\paragraph{Where do the particles come from?}
With respect to their substance our particles are indistinguishable from  their fluid or gaseous environment.
It is only their state of motion which makes the difference to the sourrounding fluid 
so that we better should talk about quasi-particles, but keep 
for brevity the term \textit{particles} in the following. 
They are characterized by their geometry,  
their specific kinetic energy, by their vectors of linear and angular momentum.

Our particles are generated as representatives of turbulent fluctuations such that 
 their energy has to be drawn from the mean flow field via so-called shear 
production\footnote{Convective turbulence 
and internal-wave breaking will be treated in another study.} 
to keep the total kinetic energy of the flow in balance. I.e.\ if there is a shear in the mean flow and non-vanishing turbulent viscosity,
then the mean flow \textit{looses} kinetic energy which reappears in form of TKE\footnote{In the following the word turbulent kinetic energy or TKE is often used.
In the given context it means actually a kinetic energy \textit{density}, i.e. energy per unit of mass. Therefore we 
measure TKE most comfortably in the units m$^2$ s$^{-2}$.}, i.e.\ as energy of vortex motions. Hereby  an exact process description of vortex \textit{generation} is not needed as long as we know \textit{how much energy} is lost and \textit{how much vorticity} is generated 
so that we can place vortices of corresponding properties  into the flow. In a way we may talk about an \textit{emergence} of  the particles. 

\paragraph{What  do the particles look like?}
Due to Helmholtz' principle of conservation of circulation, a circulation-free fluid volume 
should exhibit zero circulation over the course of time. I.e.\ particle generation cannot 
add any circulation to the volume. This implies that turbulence production can take 
place only in form of the generation of \textit{couples} of counter-rotating vortices
with zero total circulation. Consequently we specify Prandtl's fluid elements
as \textit{dipoles} where the simplest form is locally fully symmetric, which is to be understood in a statistical sense.

This assumption is central for our theory but deserves 
a comment. It is well-known that in laminar and turbulent flows wall friction 
generates velocity gradients or current shear, which is identical with circulaton induction
into the mean flow. This process is governed by RANS, more specifically by its
turbulent friction term containing  correlators like  $\langle u'\, w'\rangle$, i.e.\ by the Reynolds tensor.
This tensor is calculated further below on the basis of a particle theory. 
If a particle would have own circulation on the local level, it would add 
extra circulation to the flow, which would violate RANS. Our particles can
therefore have no own circulation. 

Thus we have the challenging theory situation that a mean flow exhibits 
turbulent friction and circulation while the turbulent vortices responsible for fluctuations and fricition 
have no circulation and are essentially frictionless, excepting the singular dissipation scale of size zero, see further below.

With respect to the specific form of vortices in our dipoles we refer to
the classical notion of \textit{vortex tubes} wherein vorticity is confined to the 
interior of a tube of smaller or larger cross section \cite[see further below and][who quote papers by Kuo \& Corrsin, 1972, and Brown \& Roshko, 1974, showing tubes as dominating characteristic structures]{pullinsaffman98}. A nice and more recent study of vortex tubes has been presented by \cite{wilczek11} who also published
on the internet a number of valuable animations of vortex ensembles in 
motion\footnote{\texttt{http://pauli.uni-muenster.de/tp/menu/forschen/}
\hspace{1cm}\texttt{ag-friedrich/mitarbeiter/wilczek-michael.html}}.
We note that, as always, the centerlines of our vortex tubes form either closed loops or they are attached to boundaries. 

%

Ordinary linear momentum of a dipole is known from classical theory in 
local approximation and can
be imagined in analogy to the motions of smoke rings. It is conserved in our 
many-particle image if each particle, generated in a locally homogeneous volume 
element, exhibits no preferred direction of motion. The linear 
momentums of two dipoles with opposite directions compensate each other.
Assuming a locally ``sufficiently high'' and \textit{even} number of
dipoles thus guarantees that the particle-induced linear momentum and circulation 
are zero and total momentum of the flow exclusively governed by the mean-flow momentum 
balance including the Reynolds stress terms.

\paragraph{How do the particles move?}
The motion of a single isolated dipole \cite[a ``naked'' particle in the terminology of many-particle physics, 
see e.g.][]{dresden93,mccomb2004} can be given by classical rules  
\cite[see][]{lamb32, albring81,saffman92, baumert05b,baumert09}. However, if the selected dipole is embedded in
a dense tangle and thus surrounded by a cloud of similar dipoles, its properties are ``screened'' by the cloud and 
thus modified: it appears to be ``dressed'' -- without violation of governing conservation laws. 
The solution is discussed in the next section.


\section{Vortices, energy, and  scales\label{linevort}}

\subsection*{Traditional vortex models}
\noindent
The  hydrodynamic literature presents a larger number of elementary analytic models for single isolated vortices 
under different conditions. Whereas the so-called potential vortex is less realistic, 
the Oseen vortex, the Rankine and the Taylor vortex, the Burgers, the Lundgren, and the Long, the Sullivan and the spherical Hill vortex are more realistic models 
of isolated vortices far from others and from boundaries \cite[for overviews see][]{lamb32,lugt79,albring81,saffman92,pullinsaffman98,davidson04}.

Excepting the potential vortex, all other vortex models are principally realistic and show first a core with radially increasing tangential 
velocity, then a saddle, and then a tail wherein the tangential velocity decreases radially down to zero. However, in our context these models are not applicable
 because they hold for conditions of isolation only. They cannot be transferred to conditions of a dense vortex tangle where the distance between 
vortex dipoles is small and the surrounding  cloud of similar vortices screens the effects from the rest. 
Therefore a new approach is needed, without violating governing conservation laws, neither on a global 
nor on the scale of the locally homogeneous and isotropic fluid volume. 

I.e.\ an ideally ``dressed'' vortex dipole moving frictionless within a tangle
of similar objects should necessarily be characterized by a finite \textit{effective} radius, 
$r$, within which all the kinetic energy, $\cal K$, and vorticity of the vortex is concentrated 
so that we may talk of an  ``energy-containing radius''. 
For simplicity, in a statistical sense $r$ and $\omega$ are taken identical 
for the two vortices forming the internally symmetric dipole.

We summarize as follows:
\begin{itemize}
\item The effective scales $\omega$ and $r$ are governed by local conservation of energy and angular momentum. 
\item 	 The radius $r$ defines a boundary within which kinetic energy and
	vorticity of a vortex are confined.
\item The effective tangential velocity of a vortex is $u=\omega \, r$ and equals the propagation velocity of the dipole. The resulting kinetic energy of the dipole is thus
${\cal K} =2\times  u^2/2 =2\times r^2\,\omega^2/2=(r\,\omega)^2$. 

\item The vortices are incompressible but deformable quasi-solid bodies which move frictionless in the vortex tangle. I.e.\ vorticity is uniformly distributed within the
cross-section of a vortex tube and has the value $2\times \omega$. 
\item At solid boundaries vortices perform frictionless roll motions. Dissipation happens exclusively $within$ the fluid at
	scale zero. (In reality the dissipation interacts with boundaries through heat and sound generation. 
\end{itemize}
These assumptions describe our image of vortices in a turbulent vortex tangle without violating
the conservation laws of kinetic energy, momentum, angular momentum and circulation. 
Compared with the above-mentioned vortex models our approach may be interpreted 
as a renormalization procedure leading to a finite kinetic energy and a finite spatial extension.

While on a first glance this procedure looks somewhat arbitrary, 
exactly this image is well established since long in physical oceanography,
which is a world of highest Reynolds numbers. 
In physical oceanography, meteorology and physical limnology the a.m. characteristic radius, $r$,
is traditionally related with the  ``energy-containing'' scale, which is a direct observable under conditions
of stable stratification and related with the so-called overturning and Thorpe scales. 
Central aspects are discussed in detail in the following section.

\subsection*{Vortex tubes and rigid-body rotation}
\noindent
In stable stratification turbulent vortices are directly recognizable in vertical profiles of density, $\rho$,
and other scalar variables, such as temperature and salinity, as ``overturning'' of density profiles. 
Within certain energetic limits they move heavier over lighter fluid and thus produce a direct imprint 
of the vortices in instantaneous depth profiles of density, $\rho(z)$. This imprint has a \textit{unique} character in
stratified flows because the buoyancy force implies that there should be a stably stratified ``reference'' state, $\bar\rho$,
a state in the absence of turbulent overturning.

\begin{figure}[h] 
 { \centering
  \includegraphics[width=8cm,height=8cm,keepaspectratio]{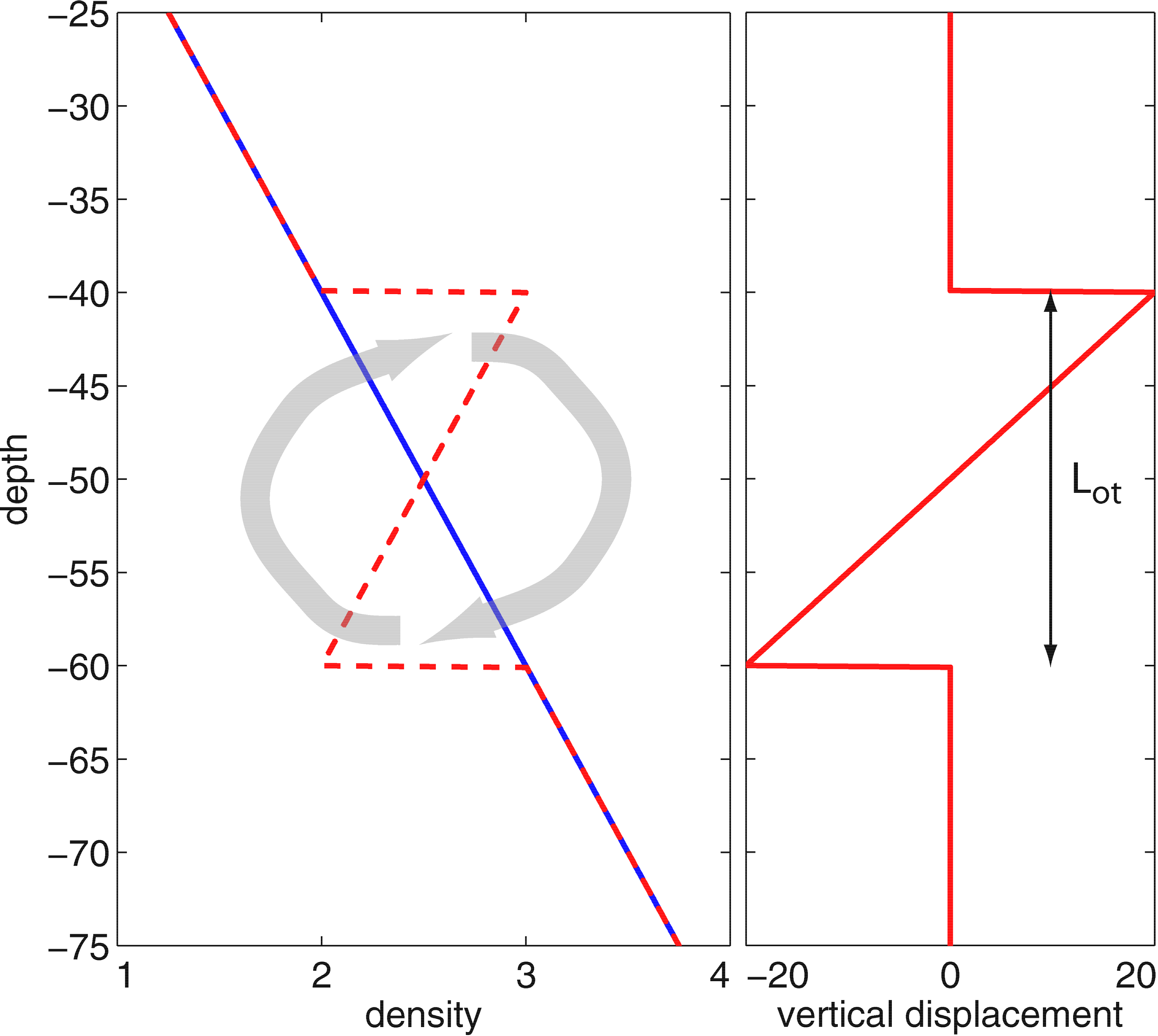}}
  \caption{Overturning and overturning scales in the sense of 
\cite[Fig.\ 6 there]{imbergerboashash86}. The cartoon depicts a
$180\deg$ overturning of a 20~m thick layer that brings relatively heavy
water up and lighter water down. Note the characteristic Z-pattern of
the vertical displacement $\zeta$. The original stable density profile is
the blue line, the unstable density in the overturn is red and dashed.
By courtesy of Dr. Hartmut Peters, Earth and Space Research, Seattle, USA.}
  \label{fig:ot1}
\end{figure}

This imprint appears typically as a Z-pattern, see Fig.\  \ref{fig:ot1}, which obviously
stands for a solid-body rotation wherein the tangential (azimutal) velocity grows linearly with the radius coordinate
up to a maximum at its outer radius and then (outside the body) drops rapidly down to zero.
We interprete a Z-pattern as a realisation of a vortex embedded 
in a dense vortex ensemble. i.e.\ as a so-called \textit{rigid  vortex} as discussed by \cite{lugt79} and further below\footnote{
The idea of a rigid vortex is compatible with fluid mechanics because Helmholtz' laws also apply here.
But the connection with continuum mechanics is not trivial because the velocity distribution in a rigid vortex
exhibits a singularity at the outer radius where it drops from a finite value down to zero. Further, the isolated
rigid vortex is not stable.}.

To our knowledge, the use of overturning deviations 
from the stable state to derive vortex radii was first introduced and explored by
\citet{thorpe77} in an analysis of temperature profiles from Loch Ness,
Scotland. Thorpe showed how the reference state can be reconstructed
from measurements and how overturns and overturning scales can be
detected. His approach is a solid basis to measure the radii (or diameters) of 
vortices in stably stratified fluids directly. 

\citet{thorpe77} described the general case of a \textit{monotonous} reference density distribution
and the overturning fluid body as \textit{frozen} for the time of the overturn. This is  justified as long as 
the length achieved through molecular diffusion during the time of rotation, 
$l_d=\sqrt{2 \, \nu_m \, t_{ot}}$, is \textit{small} compared 
with the energy-containing radius $r$ of the overturn\footnote{Here $\nu_m$ is 
molecular viscosity and $t_{ot}$ the duration of an overturn at the energy-containing 
scale $r\propto L$.}. In water this condition is always satisfied 
as $l_d$ remains here in the range of centimeters and less.

Later authors like \cite{imbergerboashash86} considered a subset of the monotonous case, 
the \textit{linear} reference-density distribution as depicted in our Fig.\  \ref{fig:ot1}. 
\cite{itsweireetal86} in their Fig.\ 1  were the first to discribe the overturning motion 
explicitely as the effect of a rigid-vortex motion \cite[cf. also][]{pullinsaffman98}.

%


\subsection*{Thorpe's method in oceanography}
\noindent
{\bf Background.} 
A short digression on the background of Thorpe's method might be in place. 
Geophysical fluid dynamics is a domain of pioneering  turbulence studies at high Reynolds numbers
because atmosphere and oceans offer the necessary conditions for free.
The first massive demonstration of measured Kolmogorov spectra 
was carried out by \cite{grantetal59} at a Reynolds number of about $Re\approx 10^8$
in a 100 m deep tidal channel\footnote{Today's DNS of turbulent flows do not
exceed $Re \approx 10^5$ and even the European superpipe CiCLOPE 
will in the best case not significantly exceed $Re\approx 10^6$ \cite[][]{ruedietal09}.}. 
They demonstrated the validity of the 5/3 law over an intervall of 
3.5 orders of magnitude (about 12 octaves). Together with the logarithmic law of the wall
and the decay of TKE in homogeneous isotropic turbulence according to $t^{-1}$ 
 \cite[or $(x/U)^{-1}$ in the wind tunnel; cf.][Chapter VII]{batchelor53}
the 5/3 law belongs until now to the most prominent universal features of high-$Re$ turbulence. 

Many observational techniques and dynamical concepts of geophysical fluid
dynamics were first developed in meteorology and later adopted in
physical oceanography and limnology. That this was not the case with turbulent
overturning is likely related to the fact that the troposphere, the most
accessible part of the atmosphere, does not show consistent mean
stratification. In contrast, oceans and lakes are almost everywhere
stably stratified outside of boundary layers. It further makes sense
that turbulent overturning was first explored in a lake where
temperature, $T$, is the sole stratifying agent. In contrast, ocean
stratification depends on temperature as well as salinity, $S$, and density
$\rho=\rho(p,T,S)$ cannot be measured with the same resolution as $T$
alone. We note in passing that real fluids are compressible such that
analyzes of overturning scales have to be based on potential density and
potential temperature.

The imprint of turbulence in vertical density profiles allows defining
and extracting overturning scales such as the Thorpe scale, $L_{Th}$
\cite[]{thorpe77}. The cartoon of Fig.\ \ref{fig:ot1} illustrates the
overturning of a $L_{ot} = 20$~m thick layer of the water column by a
single vortex in solid body rotation of diameter $L_{ot}$. 
According to our above discussion of vortex dipoles as 
quasi-particles we conclude that $L_{ot} = 2\,r$. 

The graph 
depicts the moment after a 180$\deg$ rotation that brings heavy water
up by vertical displacements with a range of $\zeta = 0 \, ... +L_{ot}$
and moves lighter water down by $\zeta = 0 \, ... -L_{ot}$. The Thorpe
scale is defined as the r.m.s.\ of $\zeta$. The linear original density
profile of the cartoon implies
\begin{equation}\label{e_lthlot}
	L_{Th} = L_{ot}/\sqrt{3} \; =\frac{2}{\sqrt{3}}\,r\;=\; 1.15\;r\; .
\end{equation}
We define $\zeta$ by the path from its original depth to a displaced
depth. That is, $\zeta$ carries the same sign as  the vertical
turbulent velocity $w'$ that 
has 
caused it.

\begin{figure}[h]
{\center
  \includegraphics[width=7.5cm,height=8.5cm,keepaspectratio]{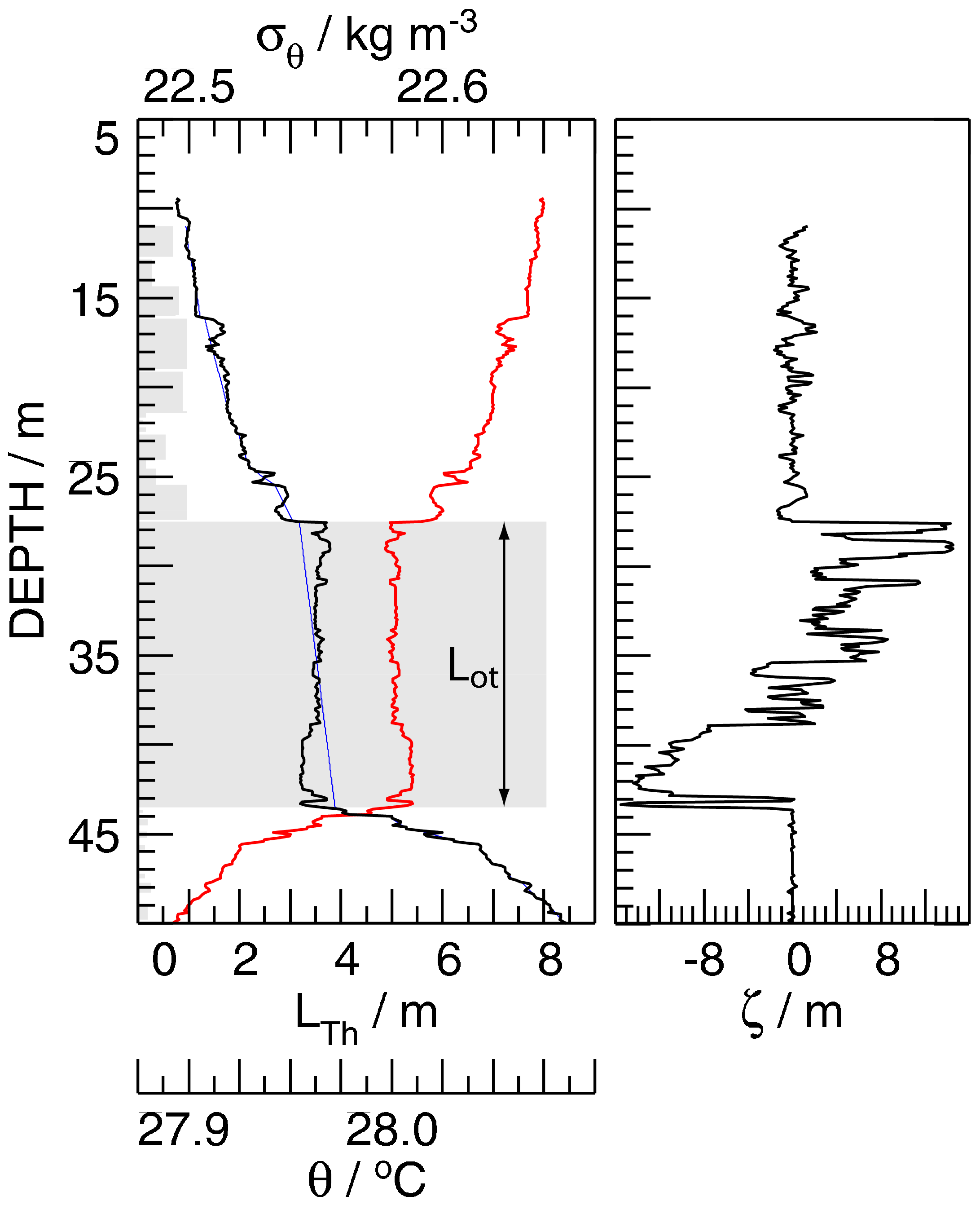}}
  \caption{A big overturn in the Pacific Equatorial Undercurrent at
$0\deg, 140\deg$W during the Tropic Heat II cruise \cite[adapted from
]{petersetal95b}: (a) potential temperature ($\Theta$, red) and
potential density ($\sigma_\Theta$, black), Thorpe-sorted
$\sigma_\Theta$ (blue),  and Thorpe scale $L_{Th}$ (shaded); 
(b) turbulent vertical displacement $\zeta$.
By courtesy of Dr. Hartmut Peters.}
  \label{fig:F157n}
\end{figure}

Fig.\ \ref{fig:ot1} may seem simplistic. It can easily be made more
realistic by adding that, in the course of the overturning, and owing to
the unstable stratification within overturns, flow instability will
occur and generate turbulence and a range of smaller scales than $L_{ot}$,
inside the big overturn. 

The ocean is full of overturns that look just like
this scenario. Fig.\ \ref{fig:F157n} depicts a big overturn in the Pacific
Equatorial Undercurrent, EUC, on the equator at $140\deg$W from
\citet{petersetal95b}. Note the Z-shape of the big overturn and its
sharp upper and lower edges.

In oceanography and limnology, Thorpe's concept of vortices overturning
parts of the water column are applied to observations as in Fig.\
\ref{fig:F157n}. Measured potential density ($\sigma_\Theta$; by convention
1000 kg~m$^{-3}$ is subtracted) or potential temperature ($\Theta$) data
points are ``Thorpe-sorted'' into monotonically rising or falling
sequences corresponding to stable density stratification. The sorted
profiles are taken as a proxy for the reference, or ``mean'' profile
that gave rise to the observed turbulence and overturning. The vertical
distance over which data points have to be moved to make the profile
monotonic is $-\zeta$. I.e.\ Thorpe-sorting ``undoes'' the overturning.
Individual overturns are defined by $\sum_i \, \zeta_i = 0$
\cite[][]{dillon82}. 

The shaded bars in Fig.\ \ref{fig:F157n}(a) show $L_{Th}$
averaged over individual overturns. This definition is highly robust. A
squared buoyancy frequency $N^2$ computed from the sorted 
$\sigma_\Theta$ or $\Theta$ is non-negative\footnote{
Here we may restrict our considerations to incompressible fluids where 
$N^2 = - g/\langle\rho\rangle d \langle\rho\rangle/ dz$ with $g$ as
gravitational acceleration, $g= 9,81$ m s$^{-2}$.}, $N^2 \ge 0$.

Turbulent length scales related to overturning scales are unaffected
by the presence of internal gravity waves (IGWs). IGWs are ubiquitous in
stratified geophysical flows and dominate velocity and scalar spectra at
vertical scales of the order of O(100~m) and smaller
 \cite[e.g.][]{petersetal95b}. The TKE accounts only for a
tiny fraction of the integral of these ``red'' spectra. Owing to this
property, length scales commonly used in laboratory experiments, such as
the \citet{ellison57} scale, $L_E =\overline{ \rho'} / \langle\partial \langle\rho\rangle /
\partial z \rangle$, are unsuitable for a characterization of turbulent length scales 
in geophysical flows because density fluctuations $\rho' = \rho - \langle\rho\rangle$ inevitably are
dominated by IGW signals so that $L_E$ can no longer be interpreted 
as a characteristic length scale of turbulent fluctuations. 
Here, angular brackets denote ensemble averages 
$\overline{\rho'}=\langle \rho'^2\rangle^{1/2}$ is an r.m.s. value.

{\bf Thorpe scale and Taylor scaling.}
The great power of Thorpe's [1977] concept can be demonstrated by
considering the so-called Taylor scaling, with $q$ as an r.m.s. measure
of turbulent velocity fluctuations. Dimensional analysis strictly gives for $l$ as
an energy-related length scale the following:
\begin{equation}\label{e_tay1}
	l \propto \, {q^3}/{\varepsilon} \;.
\end{equation}
Here $l$ is obviously related with TKE because of $\bar {\cal K}\propto q^2$,
and $\varepsilon$ is the TKE dissipation rate\footnote{We use  
the convention that the symbol ${\cal K}$ denotes the energy
of one selected dipole whereas $\bar {\cal K}$ is the local average of  
the same variable for a whole ensemble}.

\begin{figure}[h]\figurewidth{8.2cm}
\centerline{\includegraphics[width=8.2cm,height=7.5cm,keepaspectratio]{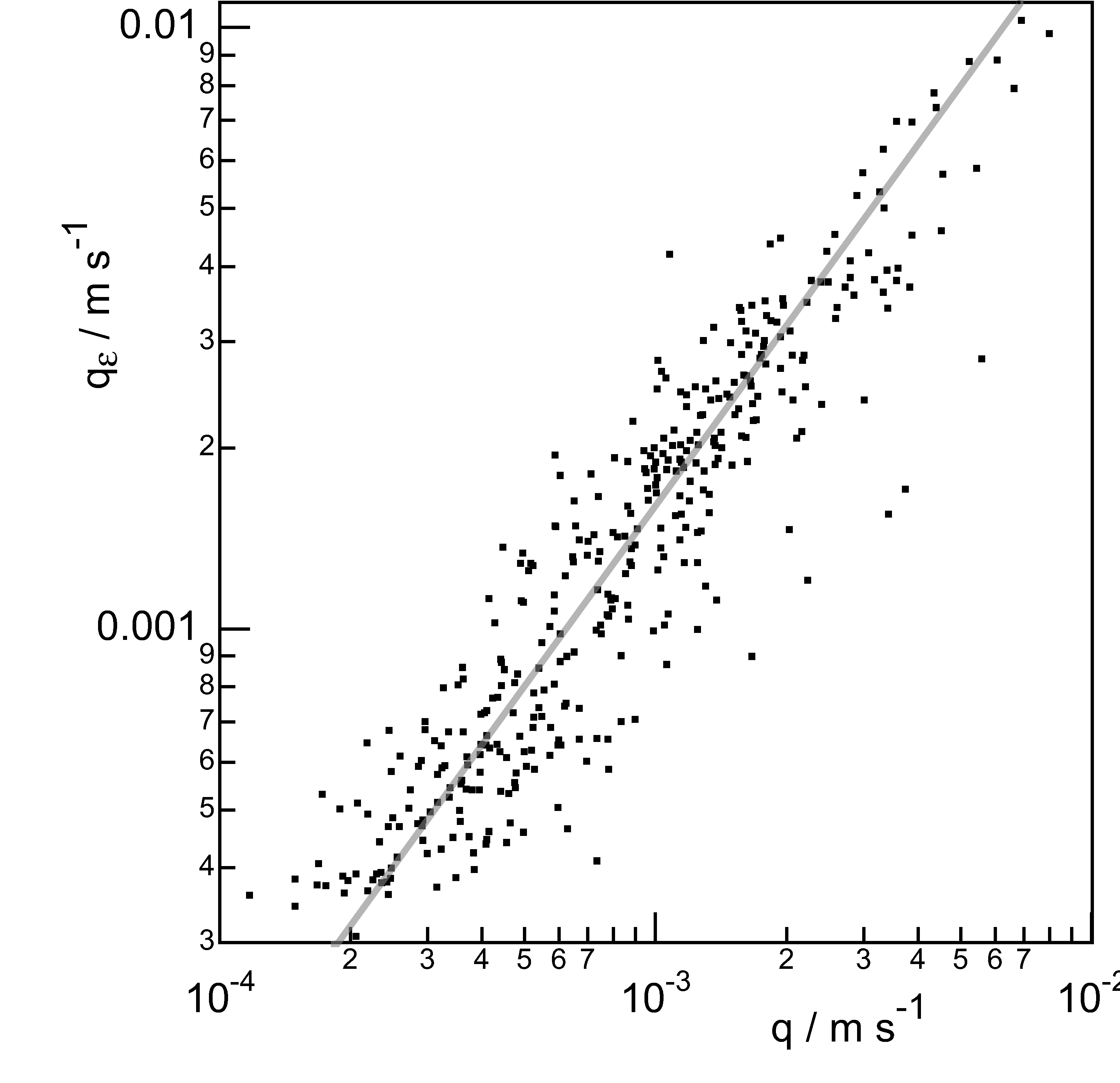}}
\caption{Turbulent velocity $q_\epsilon \!=\! (\epsilon \, l)^{1/3}$ derived from
measured $\epsilon$ via Taylor scaling, (\ref{e_tay3}), versus measured
turbulent velocity fluctuation $q$. Adapted from \citet{petersetal95b}
where $l\!=\!1.6\,L_{th}$ and where $q$ is the spectral velocity
variance at vertical wavenumbers $\ge \, 1/l$. Shown are data for
individual overturns at least 1~m thick from depths 60--350~m in the
Equatorial Undercurrent at $0\deg, \, 140\deg$W. The gray line
indicates the median ratio of $q_\epsilon$ over $q$.
 By courtesy of Dr. Hartmut Peters. \label{fig:Fqq}}
\end{figure}

If the Thorpe scale $is$ indeed a measure of the size of the energy-containing vortices
and thus related with the energy-containing scale, i.e.\ $L_{Th}\propto l$, 
then we can  write (\ref{e_tay1}) as follows,
\begin{equation}\label{e_tay2}
	\varepsilon\times L_{Th}  \propto \, {q^3}\;,
\end{equation}
or 
\begin{equation}\label{e_tay3}
	q \;\propto \, (\varepsilon \times L_{Th})^{1/3} \;.
\end{equation}
Instead of $q$ one can alternatively use $\bar {\cal K}^{1/2}$ with turbulent kinetic 
or mean vortex kinetic energy, $\bar {\cal K}$. Fig.\ \ref{fig:Fqq} shows 
measurements in the ocean demonstrating that relation (\ref{e_tay3}) is true, 
in a statistical sense. The Thorpe scale characterizes indeed the size of the energy-containing eddies. 


However, this statement needs a comment. Thorpe-sorting as a technique 
to measure the radius or diameter of energy-containing 
turbulent vortices rests on the existence, and hence works only, in stratified flows.
Under geophysical conditions this automatically implies that a certain coexistence 
of IGWs and turbulence is inescapable. The $q$ in (\ref{e_tay1}) must only
reflect TKE and must not be contaminated by the much larger IGW energy.

While the separation of waves and turbulence is beyond the scope of 
this note\footnote{The reader is referred to \citet{petersetal95b} 
and \citet{dasarolien00}.}, a workaround
is to consider spectra and to study only the ultra-violett or short-wavelengths part 
where IGW existence  is excluded through their dispersion relation.
\citet{petersetal95b} did exactly this and extracted $q$ from oceanic
 observations as the velocity variance at scales of $l$ and smaller. 
On this basis, Taylor scaling clearly emerges from their observations 
(Fig.\ \ref{fig:Fqq}).

Thorpe's [1977] concept of turbulent vortices and their imprint through vertical
overturning on density profiles allows extracting
energy-scale variables even in conditions where velocity and scalar
spectra are heavily dominated by internal gravity waves. It allows
relating energy-scale- to dissipation scale variables through the
turbulent length scale that carries Thorpe's name. These \textit{ directly
observable} energy-containing length scales stand for the 
radius or diameter of our renormalized vortices discussed 
further above.


\subsection*{Dressing a rigid-vortex tube
}
\noindent
As we learned from the considerations and examples  above, 
turbulence may be imagined as vortex tubes resembling a dense 
``local cloud'' of dipoles in chaotic motion, the cloud having zero angular and
linear momentum. So far the rigid-vortex tubes have been considered 
as somewhat arbitrary idealizations of real-world vortices.  

Translated into the language of home cooking, a snapshot of 
rigid-vortex turbulence may be 
imagined as a dense, entangled heap of hot spaghetti arrabiata. 
The spaghetti rotate around their inner centerlines and move frictionless 
within the (inviscid) sauce arrabiata. Therefore the only interactions 
between individual spaghetti occurs when they touch each other randomly. 

It arises the question whether the rigid-vortex tube and a corresponding
dipole is a \textit{stable} solution of the Euler equations. We leave it to the interested
reader to derive from these equations that an isolated rigid vortex indeed solves
these equations, but, as long as it exists, generates the following pressure head
as a consequence  of inertial 
(centrifugal) forces:
\begin{equation}\label{head}
p = p_0+ \frac{\rho }{2} \times \omega^2r^2.
\end{equation}
Here $p_0$ is the background pressure of a laminar reference flow, 
e.g.\ in the ocean the depth-depending hydrostatic pressure.
If the pressure outside the vortex would simply be $p_0$ then, 
due to the action of the outwards-directed pressure head of the vortex motion given by
the second term in (\ref{head}), the vortex would loose stability.

The stability of  quasi-rigid vortices as observed in real-world turbulence
can thus be guaranteed only by the help of a compensating force of equal strength. 
Here a concept of many-particle physics comes into play: dressing. 
As far as we will embed our initially isolated rigid vortex into
a locally homogeneous and isotropic large ensemble (``cloud'') of \textit{similar} vortices 
(more precisely: dipols made of rigid vortices), the members of the cloud generate
more or less exactly the counter pressure needed to compensate (\ref{head}) and keep
the vortex ``stable  enough''. 

According to (\ref{head}) and with $u=\omega\, r$ the pressure deviation can be written as follows,
\begin{equation}
	\delta p = p-p_0 =\frac{\rho }{2} \times u^2 = \rho \times \cal K \, ,\label{devi}
\end{equation}
where $\cal K$ is the kinetic energy density  in a vortex. Obviously the measurement
of turbulent pressure fluctuations can help estimating turbulent kinetic energy.
This corresponds to an early classical but approximate result of continuum theory
\cite[eq. (8.3.21) on p.\ 182 in][p.\ 242, last eq.]{batchelor53,hinze59}. We will not
explore the potential of (\ref{devi}) further and leave it to the interested reader. 


\section{{Turbulence and kinetic gas theory }\label{kinetic_theory}}

\subsection*{Similarities} 
\noindent
About 150 years ago, James Clerk Maxwell  presented his kinetic theory of gases.
Even if the details of molecular interaction forces in vacuum would have been known that time,
it would have not been helpful. An integration of Newtons law of motion in terms of
ordinary differential equations for each one of the billion particles was practically excluded.
But the creative use of symmetries and conservation laws radically simplified the situation and led
eventually to a closed description of the most important macroscopic properties of gases.
Main elements are the following:
\begin{itemize}
  \item[(i)] The particles in a gas are perfectly elastic \textit{points} with non-zero mass.
  \item[(ii)] They are in permanent random motion which is to be described in terms of statistical 
                 moments and sometimes called ``molecular chaos''. The scatter motions exhibit no preferred dirctions. 
  \item[(iii)] The collision results depend only on the local angular orientation of the collision partners.
  \item[(iv)] Between collisions they move uniformly and independently, without preferred direction.
  \item[(v)] Due to chemical neutrality, collisions lead only to the scatter of trajectories.
\end{itemize}  
The corresponding ``turbulent relatives'' of the above are:
\begin{enumerate}
  \item[(i')] \label{(i)}   The particles are locally symmetric vortex-dipole \textit{tubes}  
	with finite cross-sectional area, with vorticity and kinetic energy confined in the tubes, and with zero circulation.
  \item[(ii')] \label{(ii)} Their random translatory motions prefer no directions and may be termed `dipol chaos' \cite[][]{marmanis98}.
  \item[(iii')] \label{(iii)}  The result of collisions depends only on the local angular orientation of the colliding dipole elements. 
  \item[(iv')] \label{(iv)} Between collisions the particles move along complex trajectories which may be curved. 
  \item[(v')] \label{(v)} For symmetry reasons, 50\,\% of all collisions occur between two \textit{counter-rotating} dipole elements leading
	to dipole recombinations   \cite[or reconnections, like those reported in a turbulent superfluid by][]{paolettietal10} and at the end to quasi-elastic, random scatter motions 
	resembling turbulent diffusion and mixing, see left branch in Figure \ref{f:gr1}. 
\end{enumerate}  
We underline that these rules shall apply only locally in space and time. 
Of course, the prediction of the global pathway of a vortex dipole is impossible. 

\subsection*{Differences} 

\noindent
Major differences between ideal gases and turbulence are the following:
\begin{itemize}
  \item [(a)] The number of chemically neutral gas particles in a countainer is constant in time. They form a closed 
	thermodynamic system. There is no particle annihilation.
  \item[(b)] The particles of turbulence are excited energy states, their number decreases via collisions and subsequent energy dissipation. 
	They need to be replaced by shear generation of new particles if their number shall be kept more or less in balance. 
  \item[(c)] In a gas,  trajectories of point masses are simple \textit{lines} in space,  trajectories of vortex-dipole tubes produce curved \textit{areas}.
\end{itemize}
While gases evolve towards static thermodynamic equilibria, turbulence evolves either towards $dynamic$
equilibria or steady states\footnote{This neither implies  homogeneity, equifinality nor uniqueness of steady states.}, 
which may have periodic character, or turbulence dies off.

\begin{figure}[htb]
\centerline{\includegraphics[width=8cm,height=8cm,keepaspectratio]{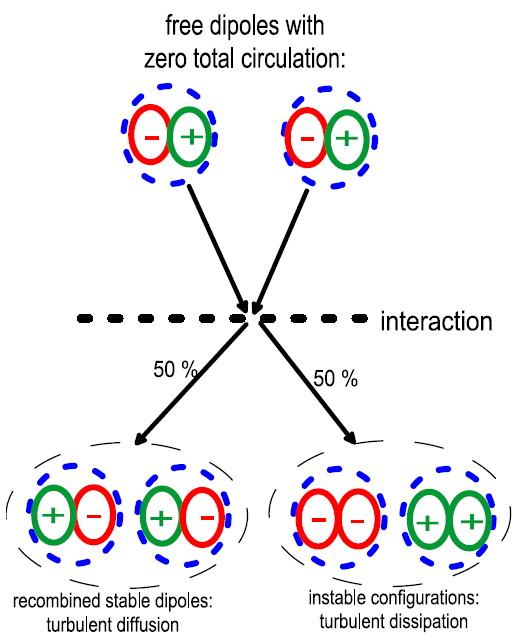}}
  \caption{Two possible collision results of two dipoles: 
the left branch is ``diffusive'' because it leads to a recombination of the dipole 
elements and scatter of trajectories which is known  as turbulent diffusion. The right branch is 
``dissipative'' because it leads to an unstable vortex configuration which decays 
``somehow'' into heat\footnote{The details of this decay process are discussed later below.}. 
For symmetry reasons both branches have identical probabilities of 0.5.
Note that the circular form of the vortex cross sections presented here are chosen 
for reasons of clarity. They hold as an statistical average picture only. 
Real vortices have elliptic or even strongly deformatted cross sections.
\label{f:gr1}}
\end{figure}

To specify this important difference we have to 
supplement our above property list for ``turbulence particles'' as follows:
\begin{enumerate}
\item[(d)]  For the same symmetry reasons like in (v') of section \ref{kinetic_theory} above, 
 the remaining 50\,\% of all dipole collisions occur between two \label{xxx}
 \textit{likewise rotating} dipole elements. Each case generates a fundamentally
 unstable vortex couple forming a slowly rotating dissipative patch with its center of mass 
more or less at rest. This patch decays through turbulent dissipation in the special form of a 
``devil's gear'' or Kolmogorov spectrum. For details see further below and the right branch 
in Fig.\ \ref{f:gr1}.
\end{enumerate}


\section{Equations of turbulent motion}

\subsection*{Brownian and turbulent motions}

\noindent
The turbulence properties (iii') and (v') of section \ref{kinetic_theory} above
establish an analogy of turbulent dipole movements with the Brownian motion
of particles suspended in a fluid at rest \cite[in the sense of][]{einstein05a}.  
In the present turbulent case it is not the kinetic theory of heat which governs 
the motions, it rather is Helmholtz' theory of dipol motions. 

 \nocite{chandrasekhar79,poeppe04,herrmann90,herrmannetal90,anderson72}

The (local) temporal path increments $\delta \vec Y_j$ of a vortex dipole $j$ may be found by 
integration  of the following Langevin equation over the time increment $\delta t$,
\begin{equation}\label{langevin3}
	\frac{d {\vec Y_j}} {dt}=\vec V_{j}\;\;,\; j = 1\dots {\cal N},
\end{equation}
where $\vec V_{j}$ is the random center-of-mass velocity of the selected dipole $j$. 

To specify the stochastic process $\vec V_{j}=\vec V_{j}(t)$ as simple as possible,
we choose a zero-mean, white-noise Gaussian process. Its strength is controlled by the locally averaged dipole properties $\bar{\cal K}$ and $\bar \omega$. 
More elaborate random processes like the
Ornstein-Uhlenbeck or the Kraichnan model \cite[][]{kraichnan68} 
and others are beyond the focus of the present paper. 

We now need to make a bigger jump over the broad river of stochastic dynamic 
systems theory where the probability density function for the solution of a dynamic 
system\footnote{A dynamic system is here understood as a set of possibly non-linear ordinary differential equations
driven at their right-hand sides by stochastic processes.} may easily be taken from
applied textbooks like those of \cite{haken78,haken1983} in form of solutions of the 
Fokker-Planck equation (FPE) corresponding to the extremely simple 
stochastic-dynamic system (\ref{langevin3}). In the present case the FPE reads as follows,
\begin{equation}\label{langevin4}
\frac{\partial {\cal N}}{\partial t} 
-\frac{\partial}{\partial \vec x} \left( \nu\;\frac{\partial {\cal N}}{\partial \vec x}\right) =0\,,
\end{equation}
where $\cal N$ is the probability density mentioned above, an equivalent of the
particle number density itself. $\nu$ characterizes the strength or intensity of the noise $V_j(t)$ in (\ref{langevin3}).
As already mentioned it is governed by the control variables ($\bar{\cal K}$, $\bar \omega$) and 
appears later as coefficient of eddy diffusivity of momentum, i.e.\ as eddy viscosity. 

In the transition from (\ref{langevin3}) to  (\ref{langevin4}) it has implicitely provided that 
all gradients exhibit sufficiently smooth and slow behavior. Those local quasi-equilibrium 
conditions are typically assumed in non-equilibrium thermodynamics and many-particle physics. 
This clearly excludes shocks and steep fronts from our considerations. 

A further comment concerns an assumption used implicitely above. It is an analogue
of the so-called Ising assumption known from the theory of magnetism:
we treat triple or higher interactions between vortex filaments as negligible\footnote{This is not fully trivial. 
Under specific conditions of Bose-Einstein condensates stable configurations consisting of one vortex 
and two anti-vortices have been observed in the laboratory, either in linear setups or
equilateral triangles \cite[][]{semanetal2009}. The latter is most symmetric and  called  a $tripole$. 
Here we assume that in our very dense ``vortex gas'' tripole-tripole interactions are controlled 
by dipole-dipole interactions of their subsets. Note that tripoles might violate the principle of zero circulation
on the level of elementary interactions needed to keep the mean-flow momentum balance in correct balance.}. 

\begin{figure}[htb]
\centerline{\includegraphics[width=7cm,height=9.0cm,keepaspectratio]{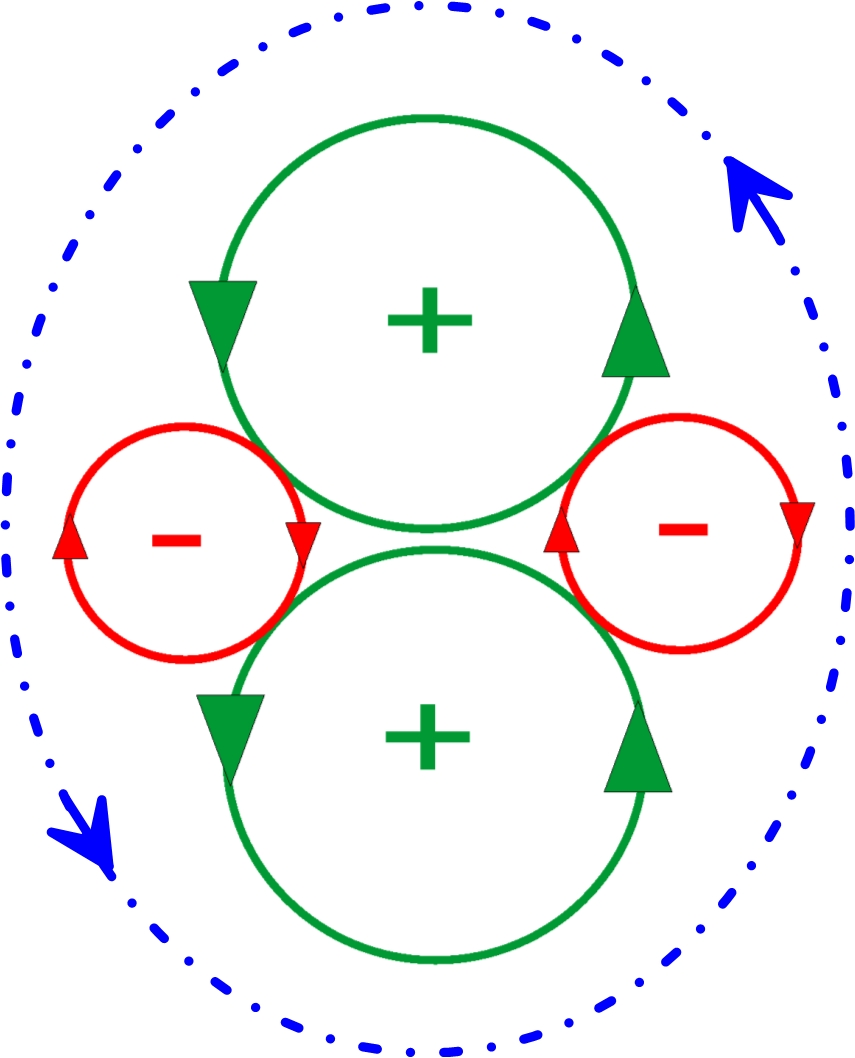}}
  \caption{Local cross section through the {first developmental stage} of an unstable pair  
of likewise rotating vortices resulting from a dipole-dipole collision (right branch in Fig.\ \ref{f:gr1}). 
The green circles represent the primary energy-containing vortices with identical radii $r$. 
Note that the green circles do \textit{not} touch each other! They are separated by
the red circles who symbolize secondary vortices in the beginning phase of a whole vortex cascade. 
The dipole cloud surrounding the above structure not only generates the necessary
pressure head to keep individual vortices stable but it also acts as a source of perturbations 
initiating roll-up instabilities and thus tertiary and higher-order vortices and eventually 
a fully developed dissipative patch, see text. 
The broken blue line and the arrows symbolize  the slow rotation of the whole patch 
around the common center of mass.  \label{f:gr2}}
\end{figure}

\subsection*{Generation and annihilation of particles}

\noindent
Property  (d) in Section \ref{kinetic_theory} above allows to supplement the 
right-hand side of equation (\ref{langevin4}) with sink and source terms, 
leading to a reaction-diffusion type equation:
\begin{equation}\label{langevin5}
\frac{\partial {\cal N}}{\partial t} 
-\frac{\partial}{\partial \vec x} \left( \nu\;\frac{\partial {\cal N}}{\partial \vec x}\right) 
=\Pi -\beta \;{\cal N}^2\;.
\end{equation}
Here $\Pi$ is the rate of quasi-particle generation. 
It is to be expressed in terms of kinetic energy per unit time and is thus 
necessarily proportional to the energy loss of the mean flow.

The second term corresponds to 
the energy-dissipation rate of TKE, $\varepsilon$, i.e. TKE conversion into heat and/or sound, with $\beta$ being a constant \cite[for details see][]{baumert09}. 
$\varepsilon$ scales with the rate of particle annihilation resulting from  collisions. 
As we learn from chemical kinetics \cite[see][for details]{haken78,haken1983}, the annihilation term is quadratic in the 
particle number, $\cal N$, because $two$ particles need to collide to generate $one$ unstable couple which is then
converted into heat and/or sound. This is discussed in the next Section.


\section{Dissipative patches}
\subsection*{Formation of vortex spectra}
\noindent
Our Fig.\  \ref{f:gr1} and the concept behind it may seem simplistic. It demonstrates the only two possible results of
a dipole collision. While the left half of the Figure shows the recombination of \textit{counter}-rotating vortices from 
counter-rotating vortices, the right half shows the the formation of a couple of \textit{likewise} rotating vortices 
from counter-rotating vortices. If isolated or naked, the likewise rotating couple revolves around a common center 
of mass and remais thus nearly at rest. This couple is known to be unstable, to form stationary dissipative patches
\cite[]{sommerfeld48}\footnote{It somehow  resembles the ``dissipative elements'' discussed by the group around 
Norbert Peters \cite[see][]{schaeferetal10}.}.

This picture with the dissipative patch almost at rest implies that dissipation is a spatially patchy phenomenon 
called   \textit{intermittency} and studied extensively by various authors \cite[for an overview see e.g.][]{frisch95}. 
We do not go into details because  intermittency is outside the focus of this study. 
We discuss instead the way \textit{how} the unstable configurations at 
the right half of Fig.\  \ref{f:gr1} and in Fig.\ \ref{f:gr2} could be transformed into a
Kolmogorov-Richardson spectrum. This problem has attracted  early attention by
\cite{taylor37} and \cite{kolmogorov41c}. The latter found on dimensional grounds that for a steady 
energy flux from large to small scales the kinetic energy spectrum as function of wavenumber
may be presented as follows\footnote{Actually, the original arguments in Kolmogorov's derivation
were more subtle, but the use of arguments in the sense of Rayleigh's method of dimensional 
analysis or, stricter, Buckingham's $\pi$ theorem, is ``strict enought'' for the present discussion.}:
\begin{equation}\label{kolmo1}
	d{\mathcal K}=\alpha _1 \,\varepsilon ^{\alpha _2}\,k^{-\alpha _3}\, dk\;,
\end{equation}
where here $k = 2\pi/\lambda $ is the wave number and $\lambda $ the wavelength. 
 $\varepsilon $ is the dissipation rate of TKE, $\cal K$.  Based on strict dimensional
arguments,  \cite{kolmogorov41c}  proved that $\alpha _2 = 2/3$ and $\alpha _3 = 5/3$, 
in excellent agreement with the famous observations by \cite{grantetal59}
in a tidal inlet with $Re \approx 10^8$ and a depth of about 100 m. 
A theoretically sound value of $\alpha_1$ was open until today and given now below.


\subsection*{Devil's gear}
\noindent
Our view of the Kolmogorov-Richardson cascade has been changed through a study by \cite{herrmann90}
who has shown that Kolmogorov's value for $\alpha_3$ corresponds numerically 
to the data  of a space-filling bearing \cite[see also][]{herrmannetal90}. 
The latter is the densest non-overlapping (Apollonian) circle packing in the plane, 
with side condition that the circles are pointwise in contact but able to rotate freely, 
without friction or slipping. One may call it a ``devil's gear''  \cite[][]{poeppe04}.

The contact condition for two different ``wheels'' with indizes 1 and 2 of the gear  reads
\begin{equation}\label{contact1}
	u = \omega_1\;r_1 = \omega_2\;r_2\; ,
\end{equation}
where $u$ is necessarily constant throughout the gear 
and governed by the energy of the decaying vortex pair as $u=\sqrt{2\,\cal K}$.
It follows that 
\begin{equation}\label{contact2}
	\omega_2 = \omega_1\;\frac{r_1}{r_2}\; ,
\end{equation}
so that for very small $r_2$ the frequency 
$\omega_2$ may become acustically relevant.

If the above gear is frictionless then the next question arises $where$ 
-- within this picture --  energy could be dissipated. Clearly, 
in a fluid with vanishing but non-zero viscosity,
 dissipation happens at
scales  where velocity gradients are high enough, here: at a scale 
of measure zero. Our dissipative patch (Fig.\  \ref{f:gr2}  shows the first
stage of its formation) is thus ``almost frictionless'' and therefore a 
Hamiltonian system, excepting scales of size zero. 

The formation of a fully developed spectrum of ``wheels'' from Fig.\ \ref{f:gr2}  deserves 
certain perturbations ``from the sides'', a condition which is guaranteed by the random 
reconnection/recombination and scatter processes sketched in
the left half of  {Fig.\ \ref{f:gr1}} and also by the incomplete mutual 
pressure  compensation of the vortices in our vortex ensemble. 

Without speculating too much 
we may expect that in a quasi-steady state patches 
like in Fig.\ \ref{f:gr2} are formed via roll-up instabilities at
the ``surface'' of the respective larger vortices. They steadily evolve 
into a fully developed gear. Its energy, dissipated at 
the smallest radii, will decrease the energy content of the 
primary (initial) vortex pair unless it is fed by mean flow.

\begin{figure}[htb]
\centerline{\includegraphics[width=5.5cm,height=3cm,keepaspectratio]{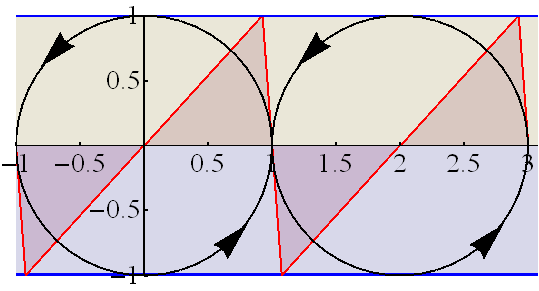}}
  \caption{Outer limits of a dissipative patch ($c.f.$ Fig.\ \ref{f:gr2}). 
The maximum wavelength is obviously equal to $\lambda_0=2\,r$. \label{f:gr3}}
\end{figure}

The outer limits of such a patch are sketched in 
Fig.\  \ref{f:gr3} for the begin of the cascade process.
The most important message which we  gain from this figure is that the $longest$
or energy-containing wavelength of the dissipative patch equals 
$\lambda_0=2\,r$. The wavelength in a  dipole is $4\,r$ and it forms no patch or spectrum.
This difference is essential. Further below we use $\lambda_0$ as
a lower integration limit for the spectral energy distribution. It is important to underline
that $\lambda_0$ labels the upper wavelength  limit in a dissipative patch.
This limit is actually not influenced by the formation details of the spectrum. 

We finally notice that the transformation of the unstable configuration
of two likewise rotating vortices deserves  time to set the greater
masses of the smaller scales into motion. This inertia effect might play a role
in highly dynamic scenarios. 

\section{Universal constants}
\noindent
In a  precursor study \cite[][]{baumert09} it has been shown that the geometric-mechanical concept 
given in the present paper allows the derivation of von-Karman's constant as 
$\kappa=1/\sqrt{2\,\pi}$.
It fueled hopes that other universal constants of 
turbulent motion might also be derived from that apparatus. 

\subsection*{Kolmogorov constant{\mathversion{bold} $\alpha_1$}\\ in the wavenumber spectrum}

\noindent
The TKE can be calculated by integrating (\ref{kolmo1}) over the dissipatve patch
in the sense sketched in Figures \ref{f:gr2} and \ref{f:gr3} yielding 
\begin{equation}\label{kolmo2}
	{\cal K} = \alpha _1 \,\varepsilon^{2/3} \, \int _{k_{0}}^{\infty} k^{-5/3}\;dk 
	=\alpha _1\,\frac{3}{2} \,\left( \frac{\varepsilon}{k_{0}}\right)^{2/3},
\end{equation}
where $k_0=2\,\pi/\lambda_0$ characterizes the lower end of the $turbulence$  spectrum 
in the wavenumber space. We loosely assign the wavenumber range 
$k=0\dots k_0$ to the mean flow which may basically be resolved in numerical models.

The dissipation rate 
$\varepsilon$ in (\ref{kolmo2}) can be expressed as follows,
\begin{equation}\label{hilfs1}
	\varepsilon = {\cal K}/\tau \,,
\end{equation}
with $\tau$ being the lifetime of a dissipative patch. 
Inserting (\ref{hilfs1}) in (\ref{kolmo2}) and rearranging gives the following:
\begin{equation}\label{kolmo13}
	\alpha_1 = \frac{2}{3}\,\left( 2\,\pi \right)^{2/3} {\cal K}^{1/3} 
	\left(\frac{\tau}{2\,r}\right)^{2/3}\,.
\end{equation}
Here we took from Fig.\ \ref{f:gr3} that the energy-containing initial or primary
 wavelength of a dissipative patch is given by $\lambda_0=2\,r$.

Now we make a local quasi-equilibrium assumption for conditions of extremely dense vortex
packing: Our marching dipoles can occupy only those places which are simultaneously 
``emptied'' from dissipative patches by decay. 
This means that the life time of a dissipative patch, $\tau ={\cal K}/\varepsilon$,
should equal the time of { ``free flight''} of a dipole over a distance 
$2\,r$ \cite[cf.][]{baumert09}:
\begin{equation}\label{quasistat}
	\tau= {\cal K}/\varepsilon = 2\,r/u\;.
\end{equation}
Here we used the scalar dipole velocity $u$,
\begin{equation}\label{dipvel}
	u=\omega\,r=\sqrt{2\,{\cal K}}.
\end{equation}
After some algebra, the dimensionless pre-factor of 
the  three-dimensional wavenumber spectrum reads as follows:
\begin{equation}\label{alpha1}
	\alpha _1\;=\;\frac{1 }{3}\,(4\,\pi )^{2/3}\;=\;1.802\,.
\end{equation}
The corresponding value of an ideal one-dimensional spectrum is one third 
of the above, i.e.\ 0.60.


\subsection*{Kolmogorov constant{\mathversion{bold} 
$\beta_1$}\\ in the frequency spectrum }

\noindent
The generalized form of the Lagrangian frequency spectrum of fluid turbulence 
is the following,
\begin{equation}\label{kolmo3}
	d{\mathcal K}=\beta _1\, \varepsilon ^{\beta _2}\,\omega ^{-\beta _3}\,d\omega \;,
\end{equation}
where $\omega $ is the angular frequency. \cite{tennekeslumley72} 
derived $\beta _2 = 1$ and $\beta _3 = 2$ in a similar fashion 
like \cite{kolmogorov41c} derived $\alpha_2$ and $\alpha_3$ 
 \cite[see also ][]{mccomb2004}.
The integration of (\ref{kolmo3}) from $\omega=\omega_0$ to $\omega=\infty$ 
gives after rearrangement and using (\ref{hilfs1}) the following,
\begin{equation}\label{kolmo4}
	\beta _1\; = \;\omega_0 \;\tau\,.
\end{equation}
Now we take from the right part of (\ref{quasistat}) $\tau=2\,r/u$,
from the left part of (\ref{dipvel}) $\omega_0 = u/r$  and insert both 
in (\ref{kolmo4}) to get finally
\begin{equation}\label{kolmo5}
	\beta _1\; = \;\omega_0 \;\tau\,=\; \frac{u}{r}\,\frac{2\,r}{u}\;=\;2\,.
\end{equation}


\subsection*{Pre-factor in the velocity\\ autocorrelation function \label{struct}}
\noindent
The spatial autocorrelation function $B(\rho)$ of fluctuating velocities is a special  second-order 
case of a structure function\footnote{For details see \S 34 in \cite{landaulifshitz_eng87}.}. 
With the abbreviation $\rho = \left| \vec \rho\right|$, it is defined as follows,
\begin{equation}\label{akf1}
	B(\rho) = \langle u_1(\vec x) \times u_2(\vec x+\vec \rho)\rangle.
\end{equation}
Here $u_1, u_2$ are the (scalar) velocity components of the flow velocity $\vec u$ 
in the direction $\vec \rho$ connecting the points $\vec x$ and $\vec x + \vec \rho$ where
the velocities $u_1$ and $u_2$ are taken respectively: 
\begin{eqnarray}\label{akf2}
	u_1 &=& \vec u(\vec x)\cdot {\vec \rho}/{\rho }\; ,\\
	u_2 &=& \vec u(\vec x+\vec \rho)\cdot {\vec \rho}/{\rho}\;.
\end{eqnarray}
The central dot denotes  the scalar product (or dot product). 
Notice that, rather than a density, $\rho$ is here a spatial distance.

In their \S 34 on p.\ 145 \cite{landaulifshitz_eng87} have shown that, based
on early results by Kolmogorov, $B(r)$ in (\ref{akf1}) may be written as follows,
\begin{equation}\label{akf3}
	B(r) = C\times (\varepsilon \, \rho) ^{2/3},
\end{equation}
where $C$ is a dimensionless numerical constant 
which is related with $\alpha_1$ from the universal
wavenumber spectrum (\ref{kolmo1}) as follows,
\begin{eqnarray}\label{akf4}
	C &=& \alpha_1 \, \frac{27}{55}\, \Gamma(1/3)\\  \label{akf5}
		&=&\frac{1 }{3}\,(4\,\pi )^{2/3}\,\frac{27}{55}\, \Gamma(1/3)\\
	&\approx& 2.37\;.\label{akf6}
\end{eqnarray}
Here $\Gamma(z)$ is the Euler gamma function.

In contrast to our derivations of $\kappa$, $\alpha_1$ and $\beta_1$ 
the value (\ref{akf6}) for $C$ should be taken with care because the 
derivation of its relation with $\alpha_1$ by Landau and Lifshitz
uses an \textit{approximation} and is thus valid for small values of the distance variable 
$\rho$ only, i.e.\ for $\ll \lambda_0 = 2\,\bar r$.


\section{Discussion}
\subsection*{Comparison with observations}

\noindent
The rounded numerical values $\kappa=0.4$, $\alpha_1=1.8$, $\alpha_1/3=0.6$, 
and $\beta_1=2$ for von Karman's and  Kolmogorov's universal constants 
predicted by our theory are situated well within the error bars of many high-$Re$ 
number observations, NSE and RG based analytical approximations, laboratory and DNS experiments. 
Based on observations, \cite{tennekeslumley72} gave the values 
$\alpha _1 = 1.62$ and $\beta _1 = 2.02$, but with greater uncertainty.

Later works have been analysed in an  important study by \cite{sreenivasan95} who
possibly gave the most comprehensive literature review of 
experimental and observational values for the number $\alpha _1/3 $ until now.
Later \cite{yeungzhou97} reported a value of $\alpha_1 = 1.62$ based on
high-resolution  DNS studies with up to 512$^3$ grid points.

\begin{figure}[htb]
\centerline{\includegraphics[width=6cm,height=5cm,keepaspectratio]{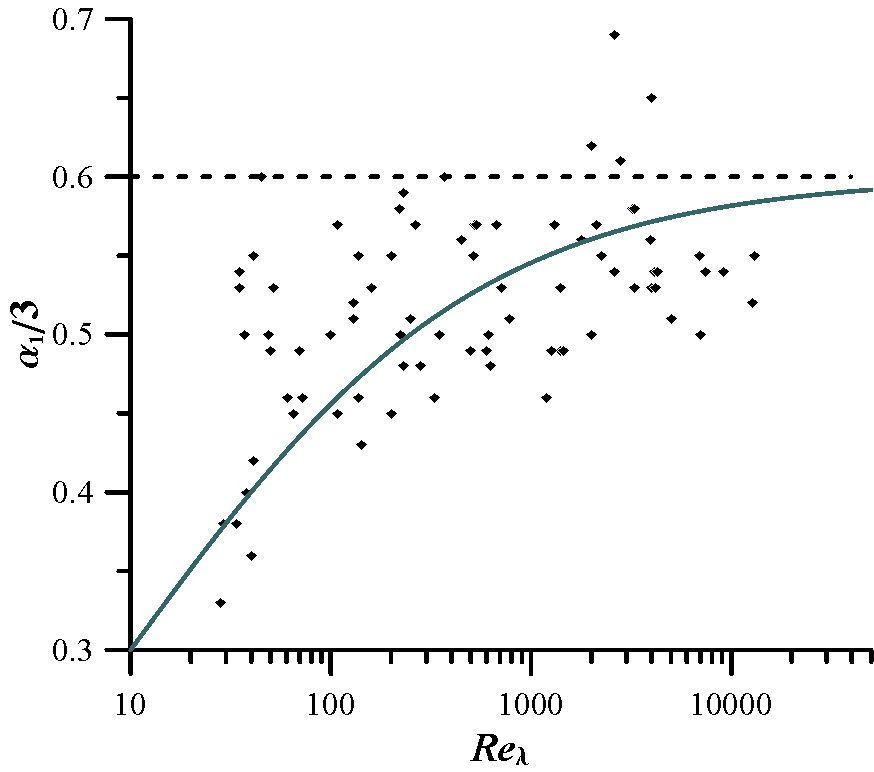}}
  \caption{Experimental and observational results  for $\alpha_1/3$
measured, collected from the literature, and analysed by \cite{sreenivasan95}. The solid green line follows our somewhat arbitrary approximation
$0.6\times {\sqrt{Re_{\lambda}}}/{\left( \sqrt{Re_*}+\sqrt{Re_{\lambda}}\right)}$ wherein
$\alpha_1/3=0.6$ is the theoretically derived  asymptotic value. Here we took $Re_{*}=10$.
A similar presentation has been chosen by \cite{sreenivasan95} for  $\alpha_1/3$ in his Fig.\ 3, 
and by \cite{yeungetal06} for $\pi \beta_1$ in their Fig.\ 2. 
\label{f:gr5}}
\end{figure}
The results of their later efforts suggest, again with DNS but based on a grid of 2048$^3$ points, 
the value $\beta_1=2.1$ \cite[][their Fig.\ 3]{yeungetal06}. 
In a recent study by \cite{donzissreenivasan10} a DNS grid of $4096^3$ has led to $\alpha_1\approx 1.58$. 

Based on their oceanic measurements  (with much higher Reynolds numbers  compared with DNS)
 \cite{liendasaro02} found that $\beta _1= 1.75 \dots 2.04$. 
They state that 
\begin{quotation}
  \dots since the present uncertainty
is comparable to that between high quality estimates of the 
Eulerian one-dimensional longitudinal Kolmogorov constant measured by many dozen investigators
over the last 50 years, large improvements in the accuracy of the estimate of 
$\beta _1$ seem unlikely.
\end{quotation}
Beginning with an initiating work by \cite{forsteretal77}, 
systematic analytical approximations using RG methods and related techniques for NSE
became further sources of estimates for the universal constants. 
E.g.\ \cite{yakhotorszag86a,yakhotorszag86b} reported $\alpha_1\approx 1.62$ whereas 
\cite{mccombwatt92} derived $\alpha_1=1.60\pm0.01$ and \cite{parkdeem03}
obtained $\alpha_1 = 1.68$. We note that these approximations are technically
extremely complex and neither unique nor part of an integrated descriptive 
concept for turbulence. 

Turbulence plays a crucial role in almost all fields of engineering, including medical applications, 
and in geophysical fluid dynamics up to climate-change studies. It plays an essential part in our everyday life. 
When we leave our house or ride our bike, when we jump into our pool -- we always are literally 
embedded in a turbulent fluid. 
Compared with most other fields of modern physics the present scatter in the 
values of the universal constants is uniquely high and therefore actually not longer acceptable. 
However, our theoretical results containing irrational and transcendental numbers suggest 
that the  universal constants of turbulence \textit{can principally not be measured} in real-world fluids. 
Or would \textit{measuring} $\pi$ be a reasonable task?

A possible solution of this dilemma is the asymptotic analysis of measurements at higher and higher 
Reynolds numbers so that an extrapolation to $Re\rightarrow\infty$ becomes feasible with some certainty.   
The   data analyses by \cite{sreenivasan95} and the report by \cite{yeungetal06} 
could serve as a methodical model. Fig.\ \ref{f:gr5} shows  a re-plot of Sreenivasan's data with $Re$ on the abscissa 
and measured $\alpha_1/3$ on the ordinate. The data exhibit a visible tendency to grow with increasing 
$Re$ to a saturation value which is statistically indistinguishable from our 0.6. 


\subsection*{Universality and fundamentality;\\ turbulence, physics, and geometry}
\noindent
Besides problems and tasks posed in engineering and geophysical fluid dynamics where 
prediction accuracy and extrapolability matter most, there is a question posed 
by natural philosophy: how are  universal constants of turbulent motion, 
fundamental physics constants like the electron's elementary 
charge  \cite[see e.g.][]{fritzsch09}, and mathematical constants like $\pi$ related with each other? 


First, the measurement accuracy of  ``turbulent numbers'' is \textit{extremely poor}. 

Second, ``turbulent constants'' characterize universal properties of a \textit{specific} (turbulent)
form of dynamic and self-similar \textit{motions} rather than more ``static'' properties
like the elementary charge. They are thus closer to the vacuum speed of light or the Hubble
constant describing the (accelerated) expansion motion of the universe. 

Third, like mathematical constants, ``turbulent numbers''  
are dimensionless whereas fundamental physics motion constants 
like the Hubble or the vacuum speed of light are given in kilometers per 
second. The latter are absolute values. ``Our'' constants characterize self-similar motions\footnote{
If we exclude from our consideration physical cosmology and arbitrarily 
chosen ratios between masses of the various atoms and molecules 
then  physics has actually only one dimensionless exception: Sommerfeld's fine-structure constant, 
with a value of about $137$. Wolfgang Pauli has long been preoccupied with the question of why, 
and Richard \cite{feynman85} even speculated about a relation with $\pi$. 
Here we have shown that $\pi$ is at least related with constants of turbulent motion.}. 

This provocates the question whether geometry is part of physics or vice versa, as discussed for
 instance by \cite{palais81}. One might argue that the
discovery of the constant angle sum in triangles was historically the first discovery
of a physical conservation law by man, made and explicitly formulated 
much earlier than the conservation laws of volume, mass etc.\footnote{
These thoughts are actually outside the scope of the present report. 
Some patience is needed as a ``theory of everything'' is not 
in sight \cite[][]{laughlin05}, and just-answered 
questions typically give birth to new conundrums.}


\subsection*{The completed image of turbulent flows}
\nocite{reynolds1895}
\noindent
For the sake of clarity we concentrate  on the most simple non-trivial situation, a spatially one-dimensional channel flow
with velocity component $U$ in horizontal ($x$)  direction, with  vertical variation along $z$, and with horizontal, $\tilde u$, and vertical velocity fluctuations, $\tilde w$. 
An example of this situation has been presented earlier 
\cite[see eq. (4.20) and (4.21) in ][where stratification is already considered through the squared buoyancy frequency, $N^2$]{baumert05a}.
The Reynolds decomposition of our flow field then reads as follows:
\begin{eqnarray}\label{perturb1}\label{U}
	U(z, t) &=& \langle U\rangle + \tilde u(z,t) \,,\\
	\label{V}
	W(z, t) &=& \langle W\rangle + \tilde w(z,t)\,.
\end{eqnarray}

{\bf Mean flow: RANS.}
One may now insert (\ref{U}, \ref{V}) into the corresponding two-dimensional Euler equation to  find together 
with mass conservation the following,
\begin{equation}\label{RANS}
	\frac{\partial \langle U\rangle}{\partial t}+\frac{\partial \langle \tilde u\,\tilde w \rangle}{\partial z}  = -\frac{\partial \langle p\rangle}{\partial x}\,,
\end{equation}
where $\langle p\rangle$ is the pressure. Equation (\ref{RANS}) is usually called a Reynolds-averaged Navier-Stokes equation (RANS).
Here we take the case of vanishing molecular viscosity. 

{\bf Turbulent mixing: downgradient  flux. }
Equation (\ref{RANS}) is not yet closed because the correlator describing a diffusive flux, $\langle \tilde u\,\tilde w \rangle$,
needs to be specified, a task which consumed substantial efforts over the last 60 years. The following quasi-linear 
flux-gradient relation is theoretically well established in many branches of many-particle physics and reads in our case as follows:
\begin{equation}\label{fluxgrad}
	 -\,\langle \tilde u\,\tilde w \rangle\;=\; \nu \, \frac{\partial }{\partial z}\,\langle U\rangle\,.
\end{equation}

{\bf Kolmogorov-Prandtl relation.}
However, also (\ref{fluxgrad}) does not yet close the problem  because now the so-called turbulent viscosity, $\nu$, needs  to be specified.
In our picture the latter is again presented in a many-particle format in analogy to Einstein's theory of Brownian motion as follows
\cite[for details see][]{baumert09}:
\begin{equation}
	\nu = L^2/\tau={\cal K}/(\pi\,\Omega)\,.
\end{equation}
Here $L=r/\kappa$ and $\tau=1/\Omega$ are locally averaged\footnote{For reasons of transparency the overbars are omitted.} space and time scales which are expressed 
within the framework of our mechanistic dipol-chaos model in terms of $\cal K$ and $\omega = 2\,\pi\,\Omega$.

{\bf Equations of turbulent motion, neutral stratification.} We skip here the technical details of derivations given in 
\cite{baumert09} and quote only the result that the variables $\cal K$ and $\Omega$ are governed by a specific 
system of nonlinear partial differential equations as follows: 
\begin{eqnarray}\label{komega-k}
	\frac{\partial\cal K}{\partial t}-\frac{\partial }{\partial z}\left( \nu\frac{\partial\cal K}{\partial z}\right)  
&=& \nu\left[ \left(\frac{\partial \langle U\rangle}{\partial z}\right)^2-\Omega^2\right],\\
\label{komega-o}
	\frac{\partial\Omega}{\partial t}-\frac{\partial }{\partial z}\left( \nu\,\frac{\partial\Omega}{\partial z}\right)   
&=& \frac{1}{\pi}\left[ \frac{1}{2} \left( \frac{\partial \langle U\rangle}{\partial z}\right)^2 - \Omega^2\right]
\end{eqnarray}
These equations resemble at least the structure of the so-called $\cal K$-$\Omega$ closure model used 
by many authors \cite[loc. cit.][]{wilcox06}. They use differing empirically gained sets of prefactors of the terms 
and  cannot identify the physical nature of $\Omega$. 
Our derivation of  (\ref{komega-k}, \ref{komega-o}) should not be confused
with a new ``scheme'' or closure of this sort because -- as we have shown -- our equations are 
founded solidly on the most simple principles of hydrodynamics and many-particle physics, 
without use of phenomenological data and, for the first time, giving even the universal constants of turbulent motion.
On the other hand, the experience of \cite{wilcox06} and colleagues shows that practical experiments and application 
needs guide the creative engineer very close to the physically correct solution, of course, without guiding further to the 
universal constants.

{\bf Turbulence spectra.}
With the dissipation rate $\varepsilon$,  
\begin{equation}\label{dissi}
	\varepsilon = {\cal K}\,\Omega/\pi\,,
\end{equation}
we may express  the spectra in the admissible turbulent 
ranges $k=k_0 \dots\infty$ and $\omega=\omega_0\dots\infty$ as follows:
\begin{eqnarray}
\label{kolmo111} 	\frac{d{\cal K}}{dk}&=&
					\alpha _1 \,\varepsilon ^{2/3}\,k^{-5/3}\,\;,\\
\label{kolmo222}	\frac{d{\mathcal K}}{d\omega}&=&
					\beta _1\, \varepsilon\;\omega ^{-2}\, \;,
\end{eqnarray}
where turbulence  begins at 
\begin{eqnarray}\label{anfk}
	k_0&=&\pi/\kappa\,L\;=\;\pi^2 \,\Omega\sqrt{2/{\cal K}}   \;,\\
\label{anfo}
	\omega_0 &=& \Omega/\kappa^2 \;=\; 2\pi\,\Omega\,,
\end{eqnarray}
respectively. Longer scales or slower motions are possibly turbulent,
but not in the sense of a fully developed spectrum. In the case of 
slow variations of the driving forces the motions outside the validity
range belong to the mean flow. Notice that the lower limits of the turbulent spectral ranges
are $dynamic$ quantities which are controlled by the dynamic variables $\cal K$ and $\Omega$.

{\bf Universal constants. }
The values of our theoretically derived universal constants of turbulent motion are
\begin{eqnarray}
\label{alpha11} 	\alpha _1&=&\frac{1 }{3}\,(4\,\pi )^{2/3}\;=\;1.802\,,\\
\label{beta11}	\beta_1 &=&2\,,\\
\label{kappa}	\kappa &=& 1/\sqrt{2\,\pi} = 0.399\,.
\end{eqnarray}

{\bf Solid boundaries.}
TKE cannot penetrate solid boundaries like walls, which is sometimes called an `adiabatic boundary condition',
in analogy to heat conduction. Assuming the wall at $z=z_*$ this means that the following condition has to be satisfied:
\begin{equation}\label{adiabatics}
\left. \frac{\partial {\cal K}}{\partial z}\right|_{z=z_*} = 0\,.
\end{equation}
This is already sufficient to solve (\ref{komega-k}, \ref{komega-o}) at solid walls and produce the
logarithmic law of the wall. A second condition for $\Omega$ is not needed due to the a.m. nonlinearities of the equations 
\cite[for details see][]{baumert05a}. 

{\bf Stable stratification.} Stable stratification may be described
by the following modification of (\ref{komega-k}):
\begin{equation}\label{komega-k-s}
	\frac{\partial\cal K}{\partial t}-\frac{\partial }{\partial z}\left( \nu\frac{\partial\cal K}{\partial z}\right)  
= \nu\left[ \left(\frac{\partial \langle U\rangle}{\partial z}\right)^2-2N^2-\Omega^2\right].
\end{equation}
This description is valid as long as $\Omega \ge N$. In the case of 
free decay and linear stratification, $\Omega$ approaches $N$ ``from above''.
Due to the dispersion relation for internal waves, slow disordered turbulent motions at $\Omega \approx N$ and slower 
can only exist as waves so that turbulence is converted into internal waves when $\Omega \rightarrow N$. 
The stratification aspects are discussed by \cite{baumertpeters04,baumertpeters05a} 
in greater detail and in relation to observations and measurements.

{\bf Limitations.}
A trivial limitation for (\ref{komega-k-s}) is $N^2 \ge 0$. For unstable or convective situations
 these equations are possibly not applicable, but we did not yet test this case. 

Equations (\ref{perturb1} -- \ref{adiabatics})   form a  closed, complete description of mean 
and turbulent motions, including their wavenumber and  frequency spectra with their universal 
spectral constants, applicable under neutrally stratified conditions. For stably stratified flows
 (\ref{komega-k})  needs to be replaced with (\ref{komega-k-s}), 
provided that there are $no$ external sources of internal-wave energy. 
This means that, due to the ubiquituous presence of internal waves in most geophysical flows,
the latter are $not$ covered by (\ref{komega-k-s}). Some modifications are 
necessary and a certain knowledge of the  sources of internal-wave energy is needed. 
The necessary modifications are not part of this report.

The difference between  (\ref{komega-k}, \ref{komega-o}) or  (\ref{komega-k-s}, 
\ref{komega-o})  on the one hand, and traditional phenomenologically based closure schemes discussed e.g.\ by \cite{wilcox06} 
on the other, is mathematically minimal but physically relevant. Those schemes use for instance 
quantities like $\varepsilon$ or $\tau$ as primary variables in balance equations although 
conservation laws for those quantities do not exist. The differences are becoming 
generally relevant and visible at solid boundaries and, in the stratified case, also within the turbulent fluid volume.

An important limitation has already been mentioned above. Further, variations in 
the mean flow field should  be relatively slow and spatial gradients not too strong so that 
we may talk about time-dependent but quasi-steady behavior with sufficiently homogeneous state 
variables on local scales and sufficiently developed Kolmogorov spectra. This excludes shock-waves from our
 considerations. But alos this limitation has not been tested yet. 

The free decay of turbulence might be such a case. Our theory predicts the decay of 
TKE to scale  with $t^{-1}$ [or with $(x/U)^{-1}$ in the wind tunnel].
This is also supported through a group-theoretical analysis by \cite{oberlack02}.
Whereas wind-tunnel  data \cite[][]{batchelor53} and selected 
free-decay measurements \cite[][]{dickeymellor80} agree surprisingly well with our theory,
in some experiments if rapid free decay the exponent is somewhat greater than unity. 
This situation is challenging and needs a further study. 

We may summarize as follows:
\begin{itemize}
  \item The present theory is based on  conservation laws and geometry only, without use of empirical data.
  \item Besides spectrally integrated parameters like turbulent viscosity, dissipation rate etc., 
%
it gives full turbulent spectra, even under non-stationary conditions.
  \item The theory predicts fundamental constants of turbulent motion.
  \item The approach is limited to slow and smooth mean flows under neutral or  stable stratification.
  \item Internal gravity waves are taken into account only so far as they are  intrinsically coupled   
		with local current shear.
\end{itemize}  
Our methodology differs from past concepts. The latter use 
specific series expansions of the Navier-Stokes equation into a 
system of partial differential equations for higher and higher moments in the perturbations, with the zero$^{th}$ 
hierarchy element being the Reynolds equation (RANS). The $k$-$\varepsilon$, the Mellor-Yamada, $k$-$\tau$, 
the $k$-$\omega$ and the many other turbulence-closure schemes populating the literature are examples 
\cite[e.g.][]{baumertetal05,wilcox06}. 

The present picture includes a perturbation step, too, but it stops at the 
zero$^{th}$ level, at RANS. Here the turbulent fluxes still remain unknown. 
We then took into account that each fluid element experiences  an 
effective total force field consisting additively of the external
mean field $\bar{\mathcal H}$ acting in the total fluid volume, and of the
 field $\tilde{\mathcal H}$ controlled by nearest neighboring vortex dipoles (``Ising assumption''),
\begin{equation}
{\mathcal H}=\bar {\mathcal H}+\tilde {\mathcal H}.
\end{equation}
Our results show that $\tilde{\mathcal H}$ is locally best described by a simple
white-noise stochastic force. This approach has successfully been used about 100 years ago by 
Langevin, Smoluchowsky and Einstein in theoretical analyses of Brownian motion. 
This assumption directly delivers Fokker-Planck equations for the expectation-value 
distribution functions of particle number and of particle 
properties $\cal K$ and $\Omega$. 

The theory applies exclusively to locally homogeneous, locally isotropic and 
weakly unsteady turbulent flows. To extreme non-stationarities and/or 
sharp spatial gradients like in shockwaves it is possibly not applicable.
As a rule, temporal changes of the mean flow should take place
on time scales sufficiently greater than $1/\Omega$ because otherwise
the Kolmogorov spectra (\ref{kolmo111}, \ref{kolmo222}) are not yet well enough established.

{\bf Coda.}
The theory completed here will hopefully not form a Procrustean bed in Saffman's  (1977) sense 
\cite[loc. cit.][p.\ 107]{davidson04} who even feared 
that ``in searching for a theory of turbulence, perhaps we are looking for 
a chimera ...''. This view has just recently been enforced by \cite{hunt2011} as follows: 
``But there are good reasons 
why the  answer to the  big  question that  Landau and Batchelor raised about whether there  is a general theory 
of  turbulence is probably {`no'}.\,''

As the above smacks of defeatism, we  pass the word to Sir Winston Churchill who demanded 
in a critical situation ``We shall never surrender!''
He surely would encourage a new generation of men and women 
to follow their own fresh ideas towards a 
{general theory of turbulence}
at asymptotically high Reynolds numbers. 
\nocite{moninyaglom2ndEd92}


\begin{acknowledgements} 
This report profited substantially from the scientific cooperation between the author 
and Hartmut Peters at Earth and Space Research in Seattle, WA, who even contributed 
Figures \ref{fig:ot1} to \ref{fig:Fqq}, and major paragraphs of Chapter \ref{linevort} of the present
report. The cooperation took place within the framework of
the Department of the Navy Grant N62909-10-1-7050 issued by the Office of Naval Research Global.
The United States Government has a royalty-free license throughout the world in all copyrightable 
material contained herein. 

The author is thankful to Eckhard Kleine who directed the attention 
to the remarkable works of Hans J. Herrmann and helped as a robust sparring partner
in turbulent debates. Special thanks are due to Stephen A. Thorpe who 
critically commented the precursor version of this report. 
Thanks are further due to David I. Benjamin,  Peter Braun, 
Eric D'Asaro, Bruno Eckhardt, Philippe Fraunier, Harald Fritzsch, 
Boris Galperin, Rupert Klein, Ren-Chieh Lien, Jim Riley, Gisbert Stoyan, 
John Simpson, J\"urgen S\"undermann,  Oleg F. Vasiliev, Michael Wilczek, 
and Sergej S. Zilitinkevich.

The author thanks particularly the  conference organizers of
\textit{Turbulent Mixing and Beyond 2011} at the Abdus Salam International
Centre of Theoretical Physics, Trieste/Italy, where the above text 
has been presented for the first time. Special thanks are due 
to Snezhana Abarzhi and Joseph J. Niemela who managed to create 
a fair, open and creative scientific atmosphere. 
\end{acknowledgements}

\bibliography{vortexdipolchaos1}

\end{document}